\documentclass[a4paper]{jpconf}
\usepackage{graphicx}
\usepackage{amssymb}

\begin{document}

\title{\textbf{A dynamical model for hierarchy and modular
organization: The trajectories en route to the attractor at the transition
to chaos}}
\author{Alberto Robledo}

\address{Instituto de F\'{\i}sica y Centro de Ciencias de la Complejidad,\\
Universidad Nacional Aut\'{o}noma de M\'{e}xico,\\
Apartado Postal 20-364, M\'{e}xico 01000 D.F., Mexico.}

\ead{robledo@fisica.unam.mx}

\begin{abstract}
We show that the full features of the dynamics towards the Feigenbaum
attractor, present in all low-dimensional maps with a unimodal leading
component, form a hierarchical construction with modular organization that
leads to a clear-cut emergent property. This well-known nonlinear model
system combines a simple and precise definition, an intricate nested
hierarchical dynamical structure, and emergence of a power-law dynamical
property absent in the exponential-law that governs the dynamics within the
modules. This classic nonlinear system is put forward as a working example
for complex collective behavior.
\end{abstract}

\section{Introduction}

Prototypical complex systems are inherently hierarchical \cite{hierarch1}-%
\cite{hierarch3}. Hierarchical systems consist of interrelated subsystems or
modules, each of which has also a hierarchical constitution. These modular
structures form levels down to a lowest one made of elementary units.
Examples of complex systems made of smaller complex systems are widespread:
astrophysical objects, social organizations, biological organisms, molecules
and atoms themselves, etc. In these kinds of systems there can be many
subsystems. For example, tissues put individual living beings together and
tissues consist of cells, a cell has nuclei and organelles and these are
made of macromolecules such as proteins. The proteins in turn are made of
many atoms, and electrons and nucleons, which are comprised of hadrons, quarks, etc,
form the atoms. In describing a hierarchy it is necessary to define how the
subsystems interact and how these interactions associate with the dynamical
processes within and between the structures. This applies to physical,
biological as well as to social complex systems.

The dynamical properties of hierarchies, within subsystems and between those
that interconnect them, are essential for the understanding of complex
systems. A purely static description is insufficient for the comprehension of the
modular organization and functioning of the entire system, and, most
importantly, for the appreciation of the properties of the system called
emergent properties, those that result in capabilities and behaviors of the
whole that are not attributable to any of the separate subsystems or
modules. Emergence `arises' from the dynamic interactions of parts within
the whole. Experience accumulated in studying complex systems have
established that the theoretical generation of dynamical hierarchies from a
formal system is not trivial and that the quest for a possible general
explanation for hierarchical structures is at least elusive. However, it is
believed that genuine emergence is strongly associated with nonlinearity,
and that a hierarchy of building blocks operating at the \textquotedblleft
Edge of Chaos\textquotedblright\ is a characteristic of complex systems.

Here we take the view that it is possible to generate hierarchical systems
with dynamical organization from a formal starting point in the form of a
simple closed-form nonlinear dynamical system. And also that nonlinear
dynamics tuned at the transition to chaos can lead to an emergent property.
We illustrate this standpoint by resorting to the properties of the simplest
well-known nonlinear iterated map, the iconic one-dimensional unimodal map,
as represented by the quadratic logistic map \cite{schuster1}, \cite%
{hilborn1}.

\begin{figure}[h]
\begin{center}
\includegraphics[width=.8\textwidth]{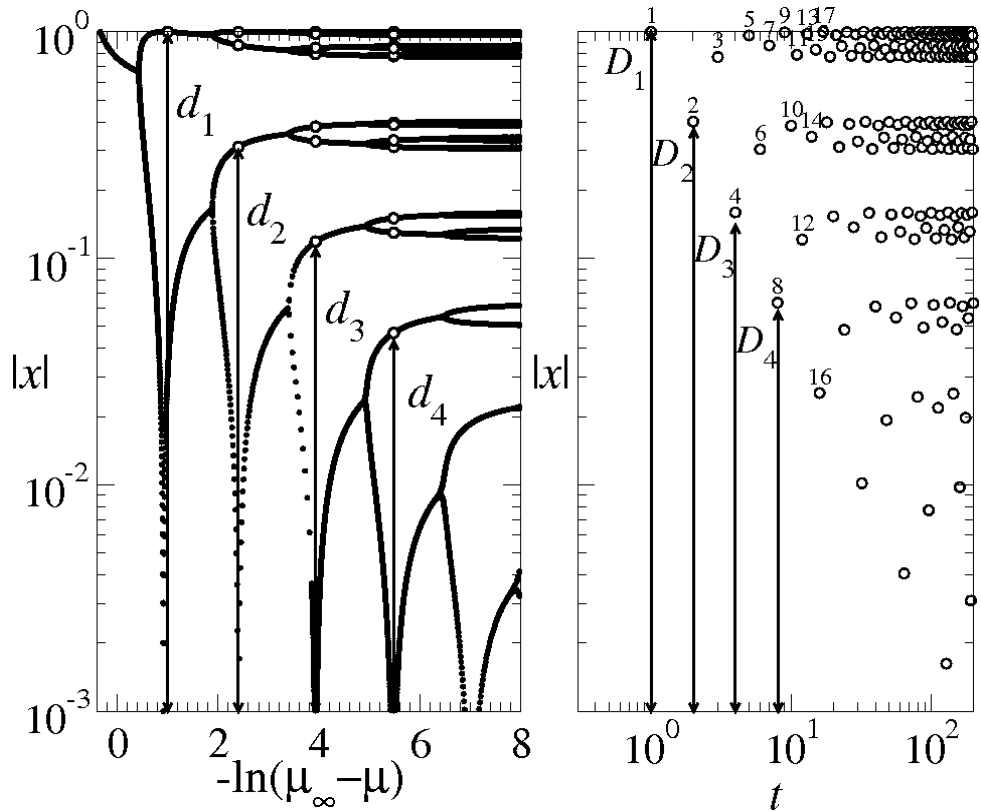}
\end{center}
\caption{{\protect\small Left panel: Absolute value of attractor positions
for the logistic map }${\protect\small f}_{\protect\mu }{\protect\small (x)}$%
{\protect\small \ in logarithmic scale as a function of }${\protect\small -}%
\ln {\protect\small (\protect\mu }_{\infty }{\protect\small -\protect\mu )}$%
{\protect\small . Right panel: Absolute value of trajectory positions for }$%
{\protect\small f}_{\protect\mu }{\protect\small (x)}${\protect\small \ at }$%
{\protect\small \protect\mu }_{\infty }${\protect\small \ with initial
condition }${\protect\small x}_{0}{\protect\small =0}${\protect\small \ in
logarithmic scale as a function of the logarithm of time }${\protect\small t}
${\protect\small , also shown by the numbers close to the circles. The
arrows indicate the equivalence between the diameters }${\protect\small d}%
_{n,0}${\protect\small \ in the left panel, and position differences }$%
{\protect\small D}_{n}${\protect\small \ with respect to }${\protect\small x}%
_{0}{\protect\small =0}${\protect\small \ in the right panel.}}
\end{figure}

Elements in our analysis are the following: i) The preimages of the
periodic attractors along the bifurcation cascade that leads to the
transition to chaos appear entrenched in a fractal hierarchical structure of
increasing complexity as period doubling develops. ii) The limiting form of
this rank structure results in an infinite number of families of
well-defined phase-space gaps in the positions of the attractor at the
transition to chaos, the so-called Feigenbaum attractor. iii) The gaps in
each of these families can be ordered with decreasing width in accordance to
power laws and are seen to appear sequentially in the dynamics generated by
uniform distributions of initial conditions. iv) The emergent power law with
log-periodic modulation associated with the rate of approach of trajectories
towards the Feigenbaum attractor is explained in terms of the progression of
gap formation. v) The relationship between the law of rate of convergence to
the Feigenbaum attractor and the inexhaustible hierarchy feature of the
preimage structure is elucidated. A detailed account of these properties is
given in Refs. \cite{robledo1}-\cite{robledo3}

\section{Brief recollection of the dynamics towards the Feigenbaum attractor}

We start by recalling that unimodal maps are dissipative due to their
one-dimensionality and consequently their dynamical properties are
associated with sets of positions called attractors as they capture at
sufficiently long times most trajectories. Other trajectories evolve towards
related sets of positions called repellors. The trajectories associated to
the period-doubling route to chaos in unimodal maps exhibit intricate
dynamical properties that follow concerted patterns. At the period-doubling
accumulation points, the transitions to chaos, periodic attractors become
multifractal before turning chaotic. We call the attractor at the transition
to chaos the Feigenbaum attractor when the Lyapunov exponent vanishes as it
changes sign \cite{schuster1}, \cite{hilborn1}. There are two sets of
properties associated with the attractors involved in these dissipative
systems: those of the dynamics inside the attractors and those of the
dynamics towards the attractors. These properties have been characterized in
detail, the organization of trajectories and of the sensitivity to initial
conditions at the attractor are described in Ref. \cite{robledo1}, while the
features of the rate of approach of an ensemble of trajectories to the
attractor has been explained in Ref. \cite{robledo2}.

\begin{figure}[h!]
\centering
\includegraphics[width=.85\textwidth]{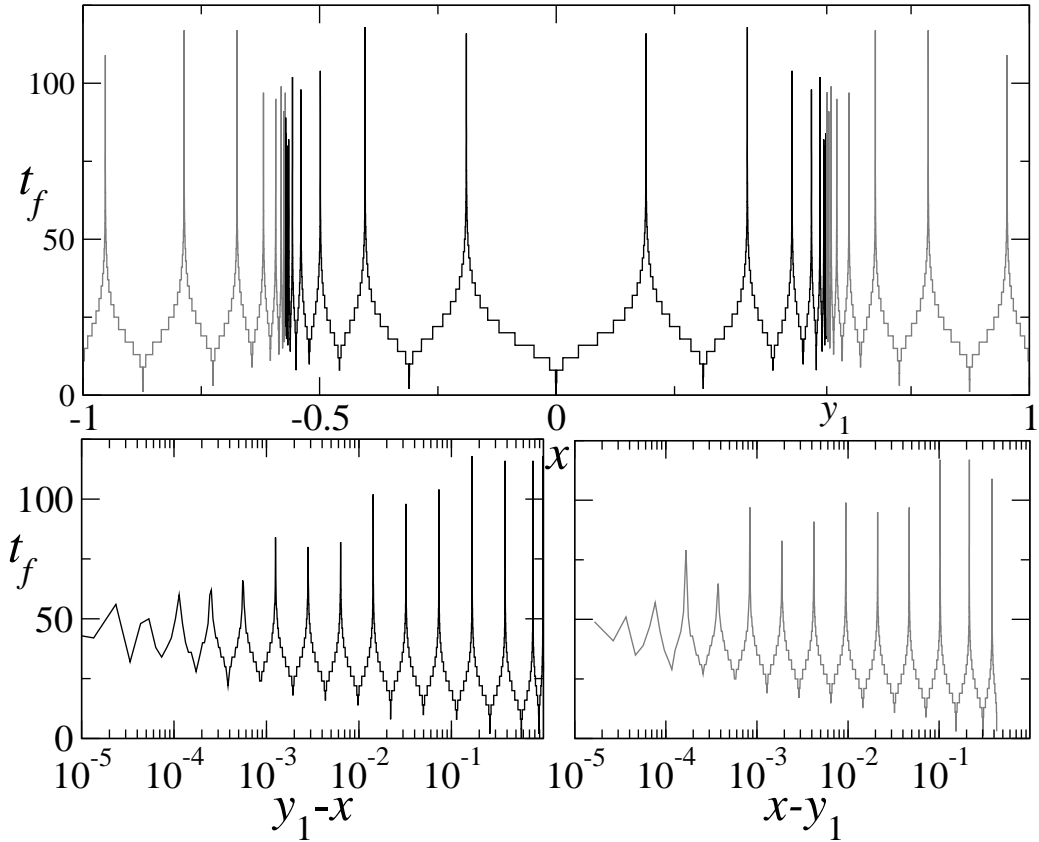}
\caption{
{\small Top panel: Time of flight }${\small t}_{f}{\small (x)}$%
{\small \ for }${\small N=2}$, the black lines
correspond to initial conditions that terminate at the attractor positions $%
{\small x=0}${\small \ and }${\small x\simeq -0.310703}${\small , while the
grey lines to trajectories ending at }${\small x=1}${\small \ and }${\small %
x\simeq 0.8734}${\small . Right (left) bottom panel: Same as top panel, but
plotted against the logarithm of }${\small x-y}_{{\small 1}}${\small \ (}$%
{\small y}_{{\small 1}}{\small -x}${\small ).  It is evident that the peaks
are arranged exponentially around the old repellor position }${\small y}_{%
{\small 1}}${\small , i.e., they appear equidistant in a logarithmic scale.}}
\label{fig2}
\end{figure}

\begin{figure}[h!]
\centering
\includegraphics[width=.85\textwidth]{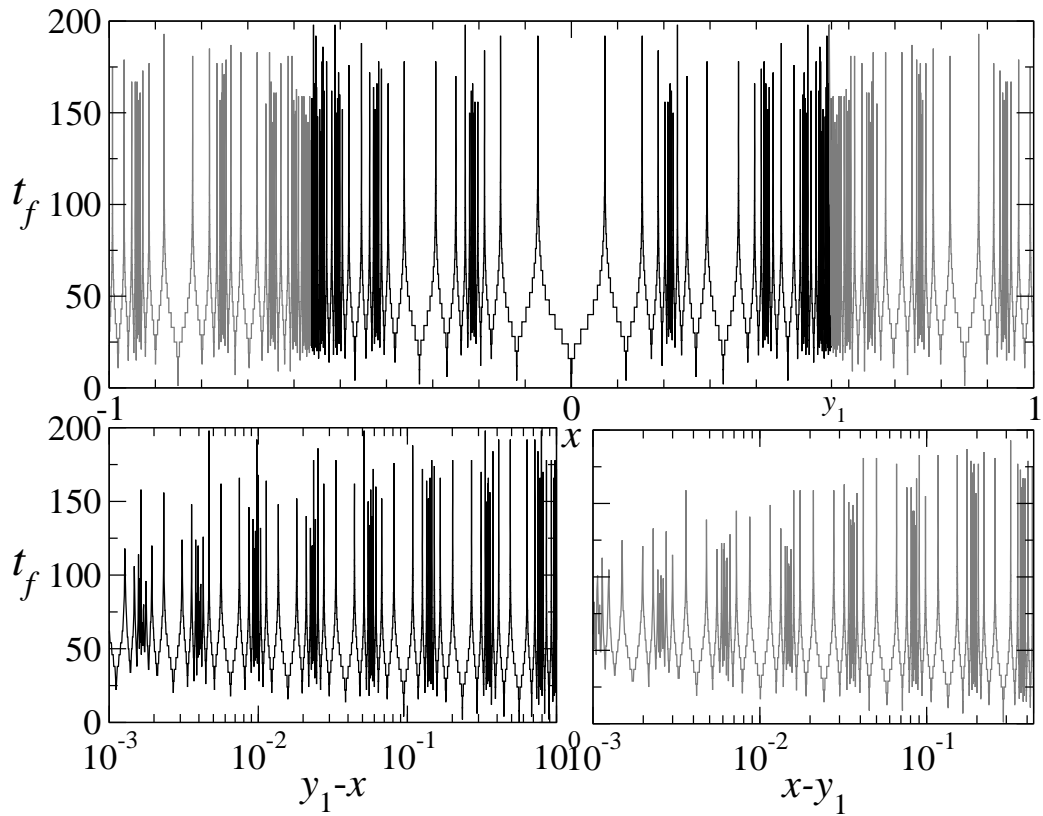}
\caption{
{\small Same as Fig. 2 for }${\small N=3}${\small . The black lines
correspond to initial conditions that terminate at any of the four attractor
positions close or equal to }${\small x=0}${\small , while the grey lines to
trajectories ending at any of the other four attractor positions close or
equal to }${\small x=1}$. {\small As the bottom panels show, in logarithmic
scale, in this case there are (infinitely) many clusters of peaks (repellor
preimages) equidistant from each other.}
}
\label{fig3}
\end{figure}

We recall now some of the basic features of the bifurcation forks that form the
period-doubling cascade sequence in unimodal maps, often illustrated by the
logistic map $f_{\mu }(x)=1-\mu x^{2}$, $-1\leq x\leq 1$, $0\leq \mu \leq 2$ 
\cite{schuster1}, \cite{hilborn1}. The dynamics towards a particular family
of periodic attractors, the superstable attractors [2], makes possible the
understanding of the rate of approach to the Feigenbaum attractor, located
at $\mu =\mu _{\infty }=1.401155189...$, and highlights the source of its
discrete scale invariant property \cite{robledo3}. The family of
trajectories associated with these attractors - called supercycles - of
periods $2^{N}$, $N=1,2,3,...$, are located along the bifurcation forks. The
positions (or phases) of the $2^{N}$- attractor are given by $x_{m}=f_{%
\overline{\mu }_{N}}^{(m)}(0)$, $m=1,2,\ldots ,2^{N}$. Notice that
infinitely many other sequences of superstable attractors appear at the
period-doubling cascades within the windows of periodic attractors for
values of $\mu >$ $\mu _{\infty }$. Associated to the $2^{N}$-attractor at $%
\mu =\overline{\mu }_{N}$ there is a $(2^{N}-1)$-repellor consisting of $%
2^{N}-1$ positions $y_{m}$, $m=1,2,\ldots ,2^{N}-1$. These positions are the
unstable solutions, $\left\vert df_{\overline{\mu }_{N}}^{(2^{n-1})}(y)/dy%
\right\vert <1$ , of $y=f_{\overline{\mu }_{N}}^{(2^{n-1})}(y)$, $%
n=1,2,\ldots ,N$. The first, $n=1$, originates at the initial
period-doubling bifurcation, the next two, $n=2$, start at the second
bifurcation, and so on, with the last group of $2^{N-1}$, $n=N$, setting out
from the $N-1$ bifurcation. See Fig. 1.

Basic to the understanding of the dynamical properties of unimodal maps is
the following far-reaching property: Time evolution at $\mu _{\infty }$ from 
$t=0$ up to $t\rightarrow \infty $ traces the period-doubling cascade
progression from $\mu =0$ up to $\mu _{\infty }$. There is an underlying
quantitative relationship between the two developments. Specifically, the
trajectory inside the Feigenbaum attractor with initial condition $x_{0}=0$,
the $2^{\infty }$-supercycle orbit, takes positions $x_{t}$ such that the
distances between appropriate pairs of them reproduce the diameters $d_{N,m}$
defined from the supercycle orbits with $\overline{\mu }_{N}<\mu _{\infty }$%
. See Fig. 1, where the absolute value of positions and
logarithmic scales are used to illustrate the equivalence. This property has
been central in obtaining rigorous results for the sensitivity to initial
conditions for the Feigenbaum attractor \cite{robledo1}. Other families of
periodic attractors share most of the properties of supercycles. We consider
explicitly the case of a map with quadratic maximum but the results are
easily extended to general nonlinearity $z>1$.

The organization of the total set of trajectories as generated by all
possible initial conditions as they flow towards a period $2^{N}$ attractor
has been determined in detail \cite{robledo3}. It was found that the paths
taken by the full set of trajectories in their way to the supercycle
attractors (or to their complementary repellors) are exceptionally
structured. We define the preimage $x^{(k)}$ of order $k$ of position $x$ to
satisfy $x=h^{(k)}(x^{(k)})$ where $h^{(k)}(x)$ is the $k$-th composition of
the map $h(x)\equiv f_{\overline{\mu }_{N}}^{(2^{N-1})}(x)$. The preimages
of the attractor of period $2^{N}$, $N=1,2,3,...$ are distributed into
different basins of attraction, one for each of the $2^{N}$ phases
(positions) that compose the cycle. When $N\geq 2$ these basins are
separated by fractal boundaries whose complexity increases with increasing $N
$. The boundaries consist of the preimages of the corresponding repellor and
their positions cluster around the $2^{N}-1$ repellor positions according to
an exponential law. As $N$ increases the structure of the basin boundaries
becomes more involved. Namely, the boundaries for the $2^{N}$ cycle develops
new features around those of the previous $2^{N-1}$cycle boundaries, with
the outcome that a hierarchical structure arises, leading to embedded
clusters of clusters of boundary positions, and so forth. The dynamics
associated to families of trajectories always displays a characteristically
concerted order in which positions are visited, which in turn reflects the
repellor preimage boundary structure of the basins of attraction. That is,
each trajectory has an initial position that is identified as a preimage of
a given order of an attractor (or repellor) position, and this trajectory
necessarily follows the steps of other trajectories with initial conditions
of lower preimage order belonging to a given chain or pathway to the
attractor (or repellor). When the period $2^{N}$ of the cycle increases the
dynamics becomes more involved with increasingly more complex stages that
reflect the hierarchical structure of preimages. See Figs. 2 and
3, and see Ref. \cite{robledo3} for details.\ The fractal
features of the boundaries between the basins of attraction of the positions
of the periodic orbits develop a structure with hierarchy, and this in turn
reflects on the properties of the trajectories. The set of trajectories
produce an ordered flow towards the attractor or towards the repellor that
reflect the ladder structure of the sub-basins that constitute the mentioned
boundaries. 

Another way by which the preimage structure described above manifests in the
dynamics of the supercycles of periods $2^{N}$ is via the successive
formation of gaps in phase space (the interval $-1\leq x\leq 1$) that
finally give rise to the attractor and repellor multifractal sets. To
observe explicitly this process we consider an ensemble of initial
conditions $x_{0}$ distributed uniformly across phase space and record their
positions at subsequent times. This set of gaps develops in time beginning
with the largest one containing the $k=0$ repellor, then followed by a set
of two gaps associated with the $k=1$ repellor, next a set of four gaps
associated with the $k=2$ repellor, and so forth. This process stops when
the order of the gaps $k$ reaches $N-1$. To facilitate a visual comparison
between the process of gap formation at $\mu =\mu _{\infty }$ and the
dynamics inside the Feigenbaum attractor---as illustrated by the trajectory
in Fig. 1 right panel---we plot in Fig. 4 the time evolution of
an ensemble composed of $10^{4}$ trajectories. We use logarithmic scales for
both $\left\vert x_{t}\right\vert $ and $t$ and then superpose on the
evolution of the ensemble the positions for the trajectory starting at $%
x_{0}=0$. It is clear from this figure that the gaps that form consecutively
all have the same width in the logarithmic scale of the plot and therefore
their actual widths decrease as a power law, the same power law followed,
for instance, by the position sequence $x_{t}=\alpha ^{-N}$, $t=2^{N}$, $%
N=0,1,2,...$, for the trajectory inside the attractor starting at $x_{0}=0$
(and where $\alpha $ is Feigenbaum's universal constant). The locations of
this specific family of consecutive gaps (the largest gaps for each value of 
$k$) advance monotonically toward the sparsest region of the multifractal
attractor located at $x=0$. Other gaps cannot be observed in the scale of
the figure due to the way the data is plotted. See Refs. \cite{robledo1}-%
\cite{robledo3} for more details.

\begin{figure}[h!]
\centering
\includegraphics[width=.8\textwidth]{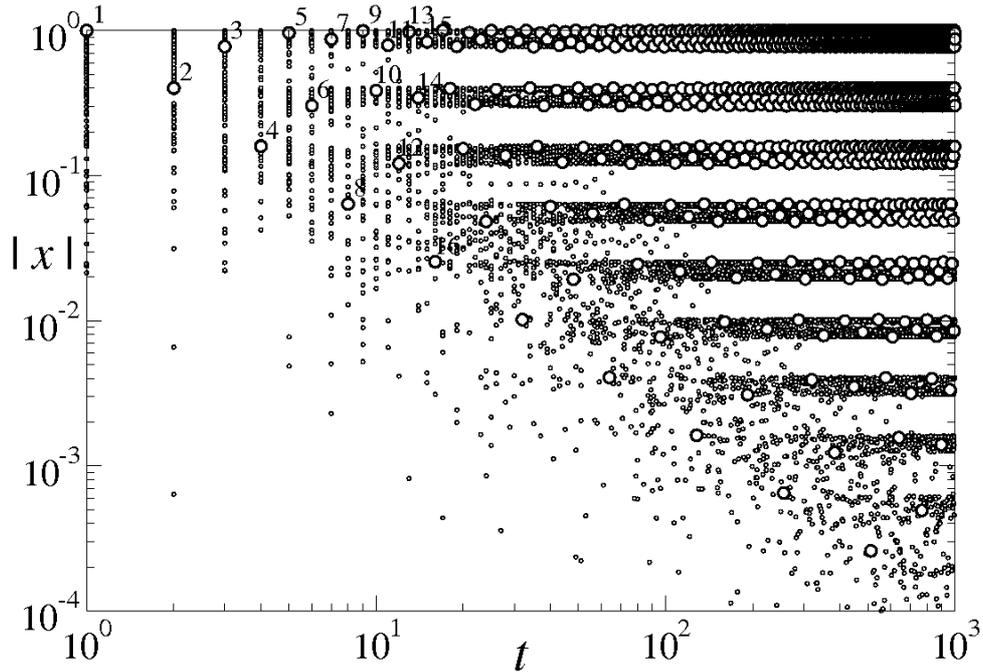}
\caption{
{\small Phase-space gap formation for }$\mu ={\small \mu }_{\infty }$%
{\small . The black dots correspond to time evolution of a uniform ensemble
of 10000 trajectories as a function of }$|x|${\small \ vs }${\small t}$%
{\small ,\ both in logarithmic scales. The open circles are the positions,
labeled by the times at which they are reached, for the trajectory inside
the Feigenbaum attractor with initial condition }${\small x}_{{\small 0}}%
{\small =0}${\small , same as right panel in Fig. 1.}
}
\label{fig4}
\end{figure}

There is straightforward quantitative way to measure the rate of convergence
of an ensemble of trajectories to the attractor and to the repellor that
consists in evaluating a single time-dependent quantity. A partition of
phase space is made of $N_{b}$ equally sized boxes or bins and a uniform
distribution of $N_{c}$ initial conditions placed, as above, along the
interval $-1\leq x\leq 1$. The number $r$ of trajectories per box is $%
r=N_{c}/N_{b}$. The quantity of interest is the number of boxes $W_{t}$ that
contain trajectories at time $t$. This is shown in Fig. 5 on
logarithmic scales for the first five supercycles of periods $2^{1}$ to $%
2^{5}$ where we can observe the following features: In all cases $W_{t}$
shows a similar initial nearly-constant plateau, and a final
well-defined decay to zero. In between these two features there are $N-1$
oscillations in the logarithmic scales of the figure. As it can be observed in the left panel of Fig.
5, the duration of the overall decay grows approximately
proportionally to the period $2^{N}$ of the supercycle. We are now in a
position to appreciate the dynamical mechanism at work behind the features
of the decay rate $W_{t}$. From our previous discussion we know that, every
time the period of a supercycle increases from $2^{N-1}$ to $2^{N}$ by a
shift in the control parameter value from $\overline{\mu }_{N-1}$ to $%
\overline{\mu }_{N}$ the preimage structure advances one stage of
complication in its hierarchy. Along with this, and in relation to the time
evolution of the ensemble of trajectories, an additional set of $2^{N}$
smaller phase-space gaps develops and also a further oscillation takes place
in the corresponding rate $W_{t}$ for finite period attractors. At $\mu
_{\infty }$ the time evolution tracks the period-doubling cascade
progression, and every time $t$ increases from $2^{N-1}$ to $2^{N}$ the flow
of trajectories undergoes equivalent passages across stages in the itinerary
through the preimage ladder structure, in the development of phase-space
gaps, and in logarithmic oscillations in $W_{t}$.

\begin{figure}[h!]
\centering
\includegraphics[width=.8\textwidth]{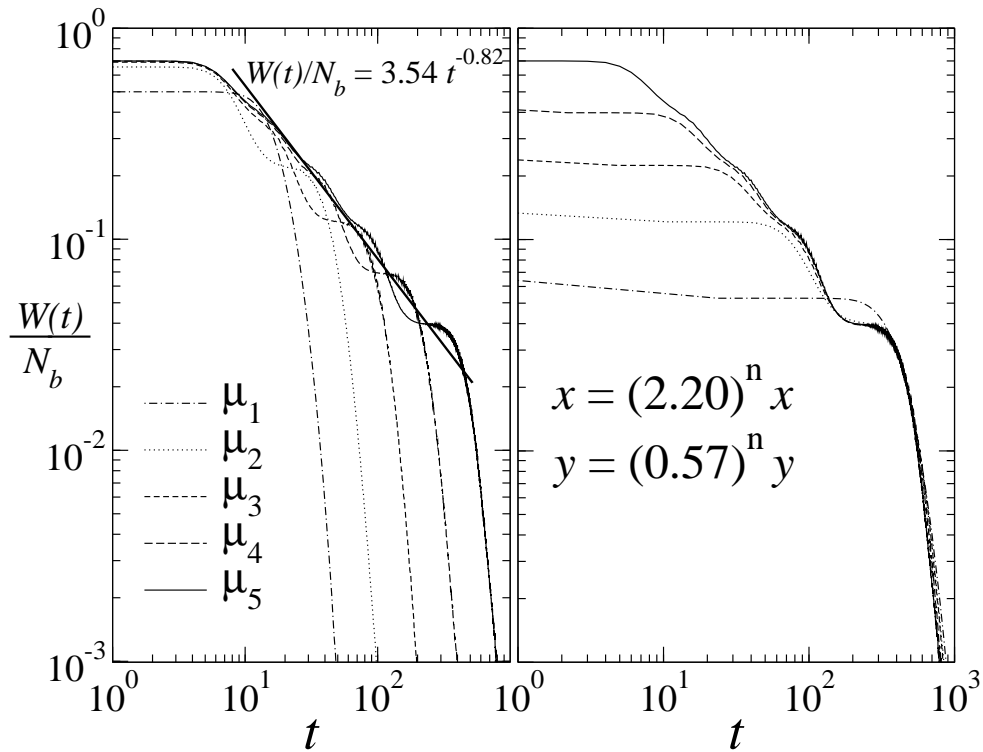}
\caption{
{\small Left panel: The rate }${\small W(t)}${\small , divided by
the number of boxes }${\small N}_{b}${\small \ employed,\ of approach to the
attractor for the supercycles of periods }${\small 2}^{{\small N}}${\small ,
}${\small N=1,2,3,4}${\small \ and }${\small 5}${\small \ in logarithmic
scales. The expression shown corresponds to the power-law decay of the
developing logarithmic oscillations. Right panel: Superposition of the five
curves for }${\small W(t)}${\small \ in the left panel via }${\small n}$%
{\small -times repeated rescaling factors shown for the horizontal }${\small %
x}${\small \ and vertical }${\small y}${\small \ axis.}
}
\label{fig5}
\end{figure}

In summary, each doubling of the period introduces additional modules or
building blocks in the hierarchy of the preimage structure, such that the
complexity of these added modules is similar to that of the total period $%
2^{N}$ system. As shown in Ref. \cite{robledo2}, each period doubling adds
also new components in the family of sequentially-formed phase space gaps,
and moreover increases in one unit the number of undulations in the
transitory log-periodic power-law decay displayed by the fraction $W_{t}$ of
ensemble trajectories still away at a given time $t$ from the attractor (and
the repellor). As a consequence of this we have obtained detailed
understanding of the mechanism by means of which the discrete scale
invariance implied by the log-periodic property \cite{robledo2} in the rate
of approach $W_{t}$ arises.

\section{The dynamics towards the Feigenbaum attractor as a dynamical
hierarchy with modular organization}

We proceed now to the identification of the features of the dynamics towards
the Feigenbaum attractor as those of a bona fide model of dynamical
hierarchy with modular organization. These are: i) Elementary degrees of
freedom and the elementary events associated with them. ii) Building blocks
and the dynamics that takes place within them and through adjacent levels of
blocks. iii) Self-similarity characterized by coarse graining and
renormalization group (RG) operations. iv) Emerging property of the entire
hierarchy absent in the embedded building blocks.

The elementary degrees of freedom at the bottom of the hierarchy are the preimages of the attractors of period $2^{2}$. These are assigned an order $k$, according to the number $k$
of map iterations they require to reach the attractor. The preimages are
also distinguished in relation to the position, $x_{1}$, \ldots , $x_{2^{2}}$%
, of the attractor they reach first. The preimages of each attractor
position appear grouped in basins with fractal boundaries. See Fig. 2. The elementary
dynamical event is the reduction of the order $k$ of a preimage by one unit.
This event is generated by a single iteration of the map for an initial
position placed in a given basin. The result is the translation of the
position to a neighboring basin of the same attractor position. The (operative) degrees of freedom higher up in the hierarchy are infinite sets of preimages of attractors of period $2^{N}$, $N>2$, for which boundary basins are replaced by clusters of boundary basins. See Fig. 3.

The modules or building blocks are clusters of clusters formed by families of boundary
basins. The boundary basins of attraction of the $2^{N}$ positions of the
periodic attractors cluster exponentially and have an alternating structure 
\cite{robledo3}. In turn, these clusters cluster exponentially themselves
with their own alternating structure. Furthermore, there are clusters of
these clusters of clusters with similar arrangements, and so on. The
dynamics associated with the building blocks consists of the flow of
preimages through and out of a cluster or block. Sets of preimages `evenly'
distributed (say one per boundary basin) across a cluster of order $N$
(generated by an attractor of period $2^{N}$) flow orderly throughout the
structure. If there is one preimage in each boundary basin of the cluster,
each map iteration produces a migration from a boundary basin to a neighboring boundary
basin in such a way that at all times there is one preimage per boundary
basin, except for the inner ones in the cluster that are gradually emptied.

The self-similar feature in the hierarchy is demonstrated by the
coarse-graining property amongst building blocks of the hierarchy that
leaves the hierarchy invariant when $\mu =\mu _{\infty }$. That is, clusters
of order $N$ can be simplified into clusters of order $N-1$. There is a
self-similar structure for the clusters of any order and a coarse graining
can be performed on clusters of order $N$ such that these can be reduced to
clusters of lower order, the basic coarse graining is to transform order $N$
into order $N-1$. Also coarse graining can be carried out effectively via
the RG functional composition and rescaling, as this transformation reduces
the order $N$ of the periodic attractor. Automatically the building-block
structure simplifies into that of the next lower order and the clusters of
boundary basins of preimages are reduced in one unit of involvedness.
Dynamically, the coarse-graining property appears as flow within a cluster
of order $N$ simplified into flow within a cluster of order $N-1$. As coarse
graining is performed in a given cluster structure of order $N$, the flow of
trajectories through it is correspondingly coarse-grained. e.g., flow out of
a cluster of clusters is simplified as flow out of a single cluster. The RG
transformation via functional composition and rescaling of the cluster flow
is displayed dynamically since by definition functional composition
establishes the dynamics of iterates, and the RG transformation $%
Rf(x)=-\alpha f(f(-x/\alpha )$ leads for $N$ finite to a trivial fixed point
that represents the simplest dynamical behavior, that of the period one
attractor. For $N\rightarrow \infty $ the RG transformation leads to the
self-similar dynamics of the non-trivial fixed point, the period-doubling
accumulation point.

There is a modular structure of embedded clusters of all orders. The
building blocks, clusters of order $N$, form well-defined sets that are
embedded into larger building blocks, sets of clusters of order $N+1$. As
the period $2^{N}$ of the attractor that generates this structure increases
the hierarchy extends and as $N$ diverges a fully self-similar structure
develops. The RG transformation for $N\rightarrow \infty $ no longer reduces
the order of the clusters and a nontrivial fixed point arises. The
trajectories consist of embedded flows within clusters of all orders. The
entire flow towards the $2^{N}$-period attractor generated by an ensemble of
initial conditions (distributed uniformly across the phase space interval of
the map) follows methodically a pattern predetermined by the hierarchical
structure of embedded clusters of the preimages.

Each module exhibits a basic kind of flow property. This is the exponential
emptying of trajectories within a cluster of order $N$. Trajectories
initiated in the boundary basins that form a cluster of order $N$ flow out
of it with an iteration time exponential law. The flow is transferred into a
cluster of order $N+1$. This flow is a dynamical module from which a
structure of flows is composed. The emerging property that appears when $%
N\rightarrow \infty $ is that there is a power-law emptying of trajectories
for the entire hierarchy. The flow of trajectories towards an attractor of
period $2^{N}$ proceeds via a sequence of stage or step flows each within a
cluster of a given order. Thus the first is through a cluster of order $1$,
then through a cluster of order $2$, etc., until the last stage is through a
cluster of order $N$. The sequence evolves in time via a power law decay
that is modulated by logarithmic oscillations. This is the emerging property
of the model.

\section{Summary and discussion}

The remarkable properties of unimodal maps already known for a long time 
\cite{schuster1}, \cite{hilborn1} have contributed significantly to the
historical development of nonlinear dynamics. The two sets of properties
that correspond to the dynamics within and towards the Feigenbaum attractor
have been studied in detail a few years ago, in the former case the
intricate structure of the sensibility to initial conditions was determined 
\cite{robledo1}, while in the latter case a surprisingly rich dynamical
organization was discovered \cite{robledo2}. Additionally, a relationship
of statistical-mechanical nature between these two types of dynamical properties
was established \cite{robledo2}.

Here we have made use of the results from these studies to identify the
features of the dynamics of an ensemble of trajectories that approach the
Feigenbaum attractor as a genuine model hierarchy where all the elements
are clearly defined and evaluated, including the observation of an emergent
property, the power-law decay with logarithmic oscillations that accompanies
the formation of the band structure of the multifractal attractor, and that
is in fact an integral part of a basic statistical-mechanical expression
that relates an entropy to a partition function \cite{robledo2}.This
statistical mechanics is of a one-parameter deformed type \cite{robledo2}. 

We have demonstrated that a dynamically-organized hierarchy with
well-defined modules arises directly from a simple formal expression with an
iteration rule without the need of any other kind of ingredients. We have
chosen a classical one-dimensional nonlinear system, a unimodal map as
exemplified by the logistic map, to demonstrate that the intricate dynamical
behavior at the transition to chaos fulfills the stipulations of such
hierarchical system. This may serve as a basic idea to build or design
different types of workable hierarchical models appropriate to different
purposes.

\textbf{Acknowledgements.} Support by DGAPA-UNAM-IN100311 and CONACyT-CB-2011-167978
(Mexican Agencies) is acknowledged.

\end{document}